\documentstyle[aps,multicol,epsf,epsfig,float]{revtex}

\def\CC{{\rm\kern.24em \vrule width.04em height1.46ex depth-.07ex
\kern-.30em C}}
\def\P{{\rm I\kern-.25em P}}
\def\RR{{\rm
         \vrule width.04em height1.58ex depth-.0ex
         \kern-.04em R}}

\def\RR{{\rm\kern.24em \vrule width.04em height1.46ex depth-.07ex
\kern-.30em R}}
\def\P{{\rm I\kern-.25em P}}
\def\RR{{\rm
         \vrule width.04em height1.58ex depth-.0ex
         \kern-.04em R}}

\newcommand{\be}{\begin{equation}}
\newcommand{\ee}{\end{equation}}
\newcommand{\bq}{\begin{eqnarray}}
\newcommand{\eq}{\end{eqnarray}}

\newcommand{\no}{\nonumber\\}

\newcommand{\diag}{\text{diag}}

\date{\today}

\begin{document}
\draft
\title{Quantum Computation with a One-Dimensional Optical Lattice}
\author{Jiannis K. Pachos\footnote{jiannis.pachos@imperial.ac.uk} and Peter L. Knight}
\address{
Blackett Laboratory, Imperial College London, Prince Consort Road, London, SW7 2BW, UK
}
\maketitle

\maketitle
\begin{abstract}
We present an economical dynamical control scheme to perform quantum computation on
a one dimensional optical lattice, where each atom encodes one qubit. The model is
based on atom tunneling transitions between neighboring sites of the lattice. They
can be activated by external laser beams resulting in a two-qubit phase gate or in
an exchange interaction. A realization of the Toffoli gate is presented which
requires only a single laser pulse and no individual atom addressing.

\end{abstract}


\begin{multicols}{2}

The rapid pace of experimental development in trapping and manipulating cold atomic
gases in optical lattices \cite{Greiner,Orzel} has inspired a series of models for
quantum information processing \cite{Bose} and quantum computation
\cite{Jaksch,Radu,Jane,Duan,Dorner,Hans,Charron}. Arrays of localized qubits
in optical lattices possess a great potential for the realization of an atom
register. Atomic trapping chips \cite{Folman,Hinds} offer ways in which atomic
arrays can be manipulated at the single-atom level \cite{Long}. In the spirit of
this investigation we present a simple model for quantum computation based on an
optical lattice with one atom per lattice site and coherent manipulation between two
different atomic ground states. Encoding the qubits in those ground states it is
possible to perform one qubit gates by Raman transitions and two qubit gates by
tunneling transitions. Adiabatic time evolutions are employed to generate a
controlled dephasing evolution or an exchange interaction between any two
neighboring atoms while preventing multiple occupancy of atoms in one site.
Alternatively, it is possible to realize the same gates by timing a fast
non-adiabatic evolution with transient population in states with two atoms per site.
This results in a significant speedup of the gates. These evolutions comprise a
universal set of gates which may be realized relatively easy experimentally. Their
main advantage derives from their simplicity and their speed compared to other
proposals \cite{Duan,Dorner}. The present scheme overcomes the problem of dephasing
along the lattice sites \cite{Hecker} by employing atoms as individual qubits.
Moreover, it offers the exciting possibility to construct multi-qubit gates such as
the Toffoli gate with minimal resources.

The proposed model consists of a one dimensional chain of atoms with two ground
states. The latter are coupled to each other with Raman transitions via an excited
level as depicted in Fig.~\ref{comb1}(a). The atoms are trapped in two parallel
in-phase optical lattices with polarization $\sigma_+$ and $\sigma_-$. These modes,
denoted by $a$ and $b$, can be generated by two counter-propagating laser beams with
parallel linear polarization vectors \cite{Mandel}. Each mode can trap one of the
atomic ground states $|g_a \rangle$ or $|g_b\rangle$. The periodicity of the lattice
is given by $\lambda/2$ where $\lambda$ is the wavelength of the laser beam from
which the standing waves are created. The loading of the optical lattice with
approximately one atom per site can be achieved by a phase transition from the
superfluid phase of a BEC to the Mott insulator phase
\cite{Jaksch1}. This is implemented by increasing the intensity of the standing wave
laser beam beyond the critical point that separates the two phases. Fock states are
obtained in each lattice site for sufficiently large intensities. The effective
Hamiltonian describing the interactions that take place is given by
\bq
&& H=-\! \sum_i \big( \!J^a_i\,a_i^\dagger a_{i+1}
     +J^b_i\,b_i^\dagger b_{i+1}
     +J^R_i \,a_i^\dagger b_i +H.c.\big)
\no \no
&& +{U_{aa} \over 2} \sum _i \left. a_i ^\dagger \right. ^2 a_i^2
+U_{ab} \sum_i a_i^\dagger a_i b_i^\dagger b_i +{U_{bb} \over 2} \sum
_i \left. b_i ^\dagger \right. ^2 b_i^2 \,\, .
\label{ham1}
\eq
The nonlinear terms $U_{aa}$ and $U_{bb}$ are produced by the collisions of atoms of
the same species with each other while $U_{ab}$ denotes interspecies collisions.
$J^a$ and $J^b$ are the couplings of the tunneling transitions and $J^R$ is the
effective coupling of the two ground states produced by a Raman transition. In
particular, consider the setup for the formation of a one dimensional lattice where
the trapping is generated by a cavity mode giving the potential
\bq
&& V_0({\bf x}) =-V_0 \sin^2 kx \exp \Big( - {2r^2 \over L^2}\Big) \,\, ,
\nonumber
\eq
where $r$ is the transverse distance from the lattice axis and $L$ is the width of
the Gaussian profile of the cavity. Within the harmonic approximation at the minima
of the lattice potential the collisional coupling is given by $U\approx4 a_s
V_0^{3/4} E_R^{1/4}/\sqrt{\lambda L} $ where $a_s$ is the scattering length of the
atomic collisions and $E_R$ is the atomic recoil energy while the tunneling rate is
given by
\bq
&& J\approx {E_R \over 2} \exp \Big( -{\pi^2 \over 4} \sqrt{V_0 \over E_R} \Big)
\Big[\sqrt{V_0 \over E_R} +\sqrt{V_0 \over E_R}^3\Big] \,\, .
\nonumber
\eq
According to \cite{Fisher,Krauth} the critical value for obtaining a phase
transition between the superfluid and the Mott phase in one dimension is given by
$U/J\approx 11.6$.

To realize a quantum computational scheme we employ the above transitions in the
following way. In the initial state the system contains one atom per lattice site,
e.g. in mode $a$. Taking the two modes to be in phase we denote by $i$ the site
corresponding to both modes. The evolution of the system is governed by Hamiltonian
(\ref{ham1}) where we shall demand individual activation of each coupling. As seen
in Fig.~\ref{comb1}(a) population can be transported from mode $a$ to mode $b$ and
back within site $i$ by performing a Raman transition between the two ground states
$|g_a\rangle$ and $|g_b\rangle$. By encoding logical $|0\rangle$ and $|1\rangle$ in
these states we can easily implement a general one qubit rotation.

\vspace{-0.4cm}
\noindent \begin{minipage}{3.38truein}
\begin{center}
\begin{figure}[th]
\centerline{
\put(15,-5){$|g_a\rangle$}
\put(70,-5){$|g_b\rangle$}
\put(46,90){$|e\rangle$}
\put(15,35){$\Omega_a$}
\put(72,35){$\Omega_b$}
\put(15,70){$\Delta$}
\put(160,-8){$i$}
\put(178,-8){$i+1$}
\put(43,-20){$(a)$}
\put(168,-20){$(b)$}
\vspace{0.3cm}
 \epsfig{file=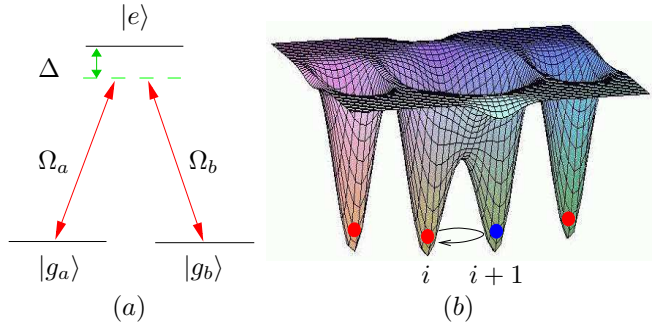,width=3.38truein}
}
\caption[contour]{\label{comb1} (a) The atomic levels with two ground states
coupled via an excited state by a Raman transition resulting to the $J^R$ coupling,
and (b) the interaction between two neighboring sites where tunneling is activated
by lowering the potential barrier between two neighboring sites.}
\end{figure}
\end{center}
\end{minipage}
\vspace{-0.7cm}

To perform a two qubit gate we proceed as follows. We assume that the potential
barriers between all sites are initially so high that no tunneling takes place
throughout the optical lattice. Consider two neighboring sites $i$ and $i+1$ in
modes $a$ and $b$ as depicted in Fig.~\ref{comb1}(b). It is possible to lower the
potential barrier between the two sites by employing an additional perpendicular
standing laser field. Its minimum should be at the position midway between sites $i$
and $i+1$ and the waist of its Gaussian profile should not be larger than two
lattice sites. This activates the tunneling between sites $i$ and $i+1$ in mode $a$
or $b$ depending on the circular polarization, $\sigma_+$ or $\sigma_-$, of the
applied laser field. In this way a hopping interaction is turned on between
neighboring atoms that results eventually to a two-qubit phase gate or to an
exchange interaction. To see how this is performed we need to consider in more
detail the structure of Hamiltonian (\ref{ham1}).

We shall take the induced tunneling couplings, $J$, to be much weaker than the
collisional ones, $U$. In this regime the tunneling transitions from one site to the
other are much weaker than the collision interactions between the atoms. We now
consider the case with only one atom per lattice site, each prepared in a given
superposition of the modes $a$ and $b$. By turning on and off the couplings $J$ such
that at all times $J\ll U$, the evolution remains in the degenerate eigenspace of
the collision terms. Indeed, the terms with coupling factors $U_{aa}$ and $U_{bb}$
are degenerate with respect to the occupation numbers $n=0$ and $n=1$. The state
with two atoms occupying one site has an energy gap from the degenerate subspace
given by $U_{aa}$ or $U_{bb}$. As a result the states with $n\geq 2$ can be
adiabatically eliminated. In addition, having a large coupling $U_{ab}$ guarantees
by the same reasoning that $n_a+n_b=1$ at all times. Simultaneous population of even
one atom per mode in the same site has an energy gap proportional to $U_{ab}$ from
the energetically lower states and hence is adiabatically avoided. As a result, the
degenerate eigenspace spanned by $|n_a=1,n_b=0\rangle$ and $|n_a=0,n_b=1\rangle$ of
every site is a well protected encoding space of the logical states $|0\rangle$ and
$|1\rangle$. Hence, quantum information processing can be performed by quantum
tunneling and by Raman transitions.
\noindent \begin{minipage}{3.38truein}
\begin{center}
\begin{figure}[ht]
\centerline{
\put(45,-25){(a)}
\put(162,-25){(b)}
\put(122,-10){$|01;10\rangle$}
\put(180,-10){$|10;01\rangle$}
\put(36,-10){$|01;01\rangle$}
\put(6,78){$|02;00\rangle$}
\put(65,78){$|00;02\rangle$}
\put(8,30){$-J^b$}
\put(75,30){$-J^b$}
\put(44,62){$U_{bb}$}
\put(122,78){$|00;11\rangle$}
\put(180,78){$|11;00\rangle$}
\put(135,35){$-J^a$}
\put(175,35){$-J^a$}
\put(108,30){$-J^b$}
\put(208,30){$-J^b$}
\put(161,62){$U_{ab}$}
 \epsfig{file=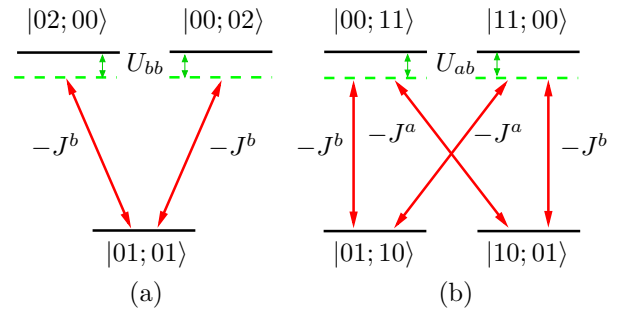,width=3.00truein}
}
\vspace{0.3cm}
\caption[contour]{\label{comb2} Effective level scheme resulting
from the tunneling and the collisional couplings. (a) The $V$-system that introduces
a phase to the state $|01;01 \rangle$ due to its coupling with large detuning to the
states $|02;00\rangle$ and $|00;02\rangle$. (b) The four level scheme with the two
Raman couplings perform transitions between the ground states $|01;10\rangle$ and
$|10;01\rangle$. }
\end{figure}
\end{center}
\end{minipage}
\vspace{-0.6cm}

It is convenient to denote the states representing the atom population of the two
sites as $|n_a^1n_b^1;n_a^2n_b^2\rangle$. Let us initially consider the case where
we lower the potential barrier between the two sites only in the $b$ mode. Then the
logical two-qubit state $|11\rangle \equiv |01;01\rangle$ couples to states with two
atoms in one site, $|02;00\rangle$ and $|00;02\rangle$, by the Hamiltonian
\[
H_1=
\left( \begin{array}{ccc}
0 & -J^b & -J^b
\\
-J^b & U_{bb} & 0
\\
-J^b & 0 & U_{bb}
       \end{array} \right) \,\, .
\]
This Hamiltonian corresponds to the level scheme presented in Fig.~\ref{comb2}(a)
that consists of a $V$-system. The effective ``Rabi frequencies" of the lasers are
both $-J^b/2$ with an equal ``detuning" $U_{bb}$. The states $|01\rangle$ and
$|10\rangle$ obtain similar phase factors by an equivalent effect. If we compensate,
in addition, for residual single qubit rotations of the form
$({J^b}^2/U_{ab})|1\rangle\langle 1|$ on both qubits we can obtain a phase change
only for state $|01;01\rangle$ which is given by $\phi=2 \int_0^T({J^b}^2/U_{ab}-
{J^b}^2/U_{bb})dt$. This is the case when $U_{bb}$ and $U_{ab}$ are much larger than
$J^b$ and adiabaticity holds. Here $T$ denotes the overall time the coupling $J^b$
is turned on. In the logical space this evolution corresponds to the two-qubit phase
gate $U=\diag(1,1,1,\exp{i\phi})$ that together with general one qubit rotations
results in universality.

It is easy to generalize this setup to realize three qubit gates like the
control-control phase ($C^2P$) and the Toffoli gate ($C^2NOT$) with, in principle,
the action of one laser pulse. Formally, the $C^2P$ gate gives a minus sign only to
the logical state $|111\rangle$ and it can generate the Toffoli gate by
$C^2NOT=({\bf 1}\otimes {\bf 1} \otimes H) C^2P ({\bf 1}\otimes {\bf 1} \otimes H)$,
where $H$ is a Hadamard gate. To generate $C^2P$  we lower the potential between
{\it three} sites of mode $b$ in such a way that tunneling interaction is activated
with coupling $J^b$ between neighboring sites and, for example, $\kappa J^b$ between
next-to-neighboring sites. Without loss of generality and for simplicity we assume
that $U_{ab}\gg U_{bb }$. For a laser pulse timed so that
$\int_0^T{J^b}^2/U_{bb}=2\pi n$ ($n$ integer) and $12 \kappa
\int_0^T{J^b}^3/U^2_{bb} = \pi$, second order perturbation theory shows that the
state $|111\rangle$ acquires a minus sign, while the states $|110\rangle$,
$|101\rangle$ and $|011\rangle$ perform a $2\pi n$ rotation and the rest of the
states are unaffected. This gives a $C^2P$ gate which can be transformed into a
Toffoli gate by applying a Hadamard gate to the third qubit. Individual laser
addressing is avoided by performing the Hadamard gate on all other qubits on the
``right" of qubit three, simultaneously.

The entangling $\sqrt{SWAP}$ gate \cite{divincenzo} can be implemented by
engineering an exchange interaction between two neighboring sites. Let us consider
the evolution of the degenerate states $| 10 ;01\rangle$ and $| 01 ; 10\rangle$ when
both of the tunneling rates $J^a$ and $J^b$ are activated. In the basis
$|00;11\rangle$, $| 10 ;01\rangle$, $| 01 ; 10
\rangle$ and $|11;00\rangle$ the evolution is dominated by the Hamiltonian
\[
H_2=
\left( \begin{array}{cccc}
U_{ab} & -J^a & -J^b & 0
\\
-J^a & 0 & 0 & -J^b
\\
-J^b & 0 & 0 & -J^a
\\
0 & -J^b & -J^a & U_{ab}
       \end{array} \right) \,\, .
\]
At the same time the phase evolutions given by $H_1$ occur to both modes $a$ and
$b$. Adiabatically eliminating the transitions outside the degenerate subspace
provides an evolution similar to a Raman transition with the equivalent ``Rabi
frequencies" being $-J^a$ and $-J^b$ and with a ``detuning" $U_{ab}$. The degenerate
eigenspaces together with their couplings are depicted in the level scheme of
Fig.~\ref{comb2}(b). If the initial population is in the logical space and the
tunneling rates are weak, then the adiabatic evolutions remains in this space
resulting effectively in the Hamiltonian
\be
H_{\text{eff}} = - I (|10\rangle \langle 01| + |01\rangle \langle 10|)
\ee
where $I=2J^aJ^b/U_{ab}$. The phase evolutions can be factorized out by choosing the
couplings to satisfy the condition ${J^a}^2 /U_{aa}+{J^b}^2/U_{bb}=
({J^a}^2+{J^b}^2)/ U_{ab}$. In addition, one can compensate for residual single
qubit rotations of both qubits of the form $({J^a}^2/U_{aa}-
{J^b}^2/U_{bb})(|0\rangle\langle 0|-|1\rangle \langle1|)$  by applying properly
tuned lasers.
\noindent
\begin{minipage}{3.38truein}
\begin{center}
\begin{figure}[ht]
\centerline{
\put(40,-10){$A=0$}
\put(117,-10){$A={\pi \over 2 }$}
\put(195,-10){$A=\pi$}
 \epsfig{file=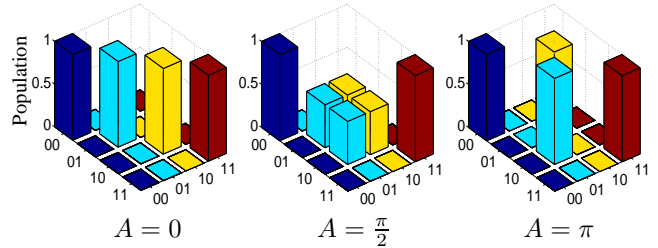,width=3.38truein}
}
\vspace{0.3cm}
\caption[contour]{\label{evolution} Evolution of the density matrix elements
$|\rho_{ij}|^2$ that comprise the two qubit space $\{|ij\rangle; \text{for
}i,j=0,1\}$ for different values of the action $A=\int I dt$ that activates the
exchange interaction.}
\end{figure}
\end{center}
\end{minipage}
\vspace{-0.7cm}

In order to perform the $\sqrt{SWAP}$ gate we now turn on the coupling $I$, via the
tunnel $J$ couplings for a certain time so that the action $A=\int I dt $ equals
$\pi/2$. For $A=\pi$ we obtain the $SWAP$ gate. The effective evolution of two
qubits produced by $H_{\text{eff}}$ resulting in the exchange interaction is given
in Fig.~\ref{evolution}. The simulation was carried out with the full Hamiltonian
(\ref{ham1}) for a ratio of the couplings $J/U=10^{-2}$, which is well within the
adiabatic weak coupling limit.

\noindent
\begin{minipage}{3.38truein}
\begin{center}
\begin{figure}[ht]
\centerline{
 \epsfig{file=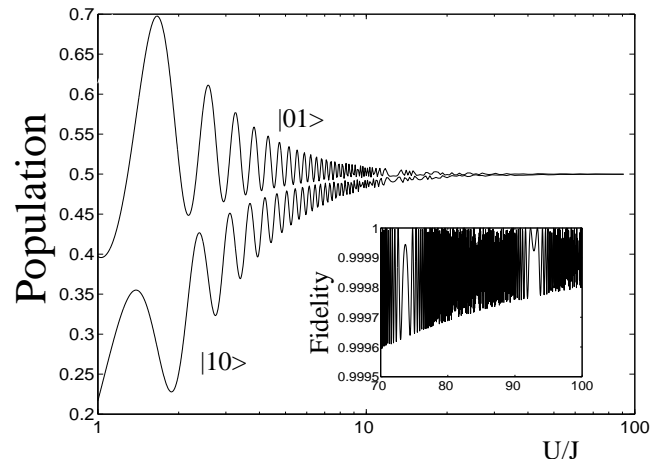,width=3.38truein}
}
\vspace{0.3cm}
\caption[contour]{\label{Umodulations5} The population of the states $|01\rangle$
(upper) and $|10\rangle$ (lower) when the tunneling interactions are activated with
$A=\pi/2$ on the initial state $|01\rangle$ as a function of $U/J$. The inset
presents the fidelity of the exchange interaction for producing a maximally
entangled state that deviates from unity by $4 \cdot 10^{-4}$ or less.}
\end{figure}
\end{center}
\end{minipage}
\vspace{-0.7cm}

The condition of weak couplings $J$ with respect to the $U$'s is necessary in order
to adiabatically eliminate any evolution of the qubit states that would result in a
population of two atoms in one lattice site. Indeed, by employing the effective
exchange interaction to generate a maximally entangled state the fidelity of our
procedure is $4 \cdot 10^{-4}$ or less away from unity. In Fig.~\ref{Umodulations5}
we see this fidelity for different ratios of $U$ and $J$. For $U\sim 10^2 J$ it
exhibits strong oscillations due to the large collisional couplings, but with a very
small amplitude indicating the success of the adiabatic transition. This is paid for
by obtaining slow gates. For currently measured collisional couplings
\cite{Mandel} of the order of $1\,$kHz a fidelity deviation
from one less than $10^{-3}$ can be achieved if the duration of a macroscopic gate
(e.g. a $\pi$-phase gate) is $T\sim 100\,$ms. For these parameters the errors in the
tunneling rate due to fluctuations of the laser intensity of the order of $10^{-3}$
contribute an acceptable gate error of the order of $10^{-3}$ or lower.

Alternatively, fast evolutions can be achieved by strong tunneling couplings. During
these operations population is transferred into multi-occupancy states of the sites,
but completely returns back to the logical space at certain times. Hence, by
carefully controlling the timing of the tunneling procedure we obtain much faster
gates. For example, the realization of a fast controlled phase-gate employs the same
level scheme as in the previous. By timing the evolution so that $T_n=2\pi n
/\sqrt{U_{bb}^2 +8{J^b}^2}$ we obtain the two qubit phase-gate with $\phi=\pi n\Big[
1+(U_{bb}-2U_{ab})/\sqrt{U_{bb}^2 +8{J^b}^2}\Big]$. The condition
$J^b=\sqrt{U_{ab}^2-m^2/n^2 U_{bb}^2}/(2\sqrt{2m^2/n^2-1})$, ($m$, $n$ are positive
integers with $m> n/\sqrt{2}$) guarantees that any possible transient population of
two atoms per site is eliminated at the end of the gate.

Furthermore, it is possible to time the exchange interaction such that even for
strong tunneling couplings there is zero population of states with two atoms per
site regardless of their internal states (see Fig. \ref{Umodulations5}). Consider
for simplicity the couplings $J^a=J^b=J$ and $U_{aa}=U_{bb}=U$. In this case the
time interval for a completion of this evolution is given by $T_n=2 \pi
n/\sqrt{U_{ab}^2+16 J^2}$ where we require $J=\sqrt{U^2-m^2/n^2 U_{ab}^2}/
(2\sqrt{4m^2/n^2-2})$. The condition for no-phase evolution of the states
$|00\rangle$ and $|11\rangle$ is $U=2U_{ab}$ if $m$ is an even integer. The
resulting two-qubit gate $G$ is obtained at times $T_n$ and is given by $G(T_n) =
(|00\rangle\langle00| + |11\rangle\langle11|) + [1+(-1)^n e^{-i/2 UT_n
}]/2(|01\rangle \langle01| + |10\rangle \langle 10|) + [-1+(-1)^n e^{-i/2 UT_n
}]/2(|01\rangle \langle10| -H.c.)$. This is in general a non-trivial gate that
results together with any one-qubit gate in universality. These fast evolutions
should be performed with couplings $J$ much smaller than the band gap of the trapped
atoms in order to avoid population of higher vibrational modes of the optical
trapping potential. In addition, the laser amplitude has to be stabilized in a more
precise fashion than in the previously described adiabatic evolutions in order to
achieve similar requirements in the gate precision. From such procedures we can
achieve gate operation times of a few ms.

In conclusion, we have presented a simple scheme for performing two qubit gates
based on adiabatic passages with tunneling interactions, which can effectively
reproduce a phase gate or an exchange interaction. In addition, we have seen how
three-qubit gates as the Toffoli gate can be easily generated without evoking
individual atom laser addressing. The gate operation time can be greatly reduced by
increasing the tunneling coupling and precise timing of the non-adiabatic evolution.
In contrast to previous models for quantum computation on an optical lattice, the
gates proposed here act locally between two neighboring qubits, while the rest of
the lattice is non-interacting. In this way we are immune against the major
experimental problem of dephasing of the lattice sites due to longitudinal
irregularities of the lattice potential. In the future, it would be of much interest
to study the possibility of performing the existing quantum algorithms in terms of
common one-qubit gates and multi-qubit gates. As we have seen those are easy to
realize within the present scheme, while individual one-qubit gates are
experimentally much more difficult to implement.
\\
\vspace{-0.3cm}

{\em Acknowledgments.}  We thank E. A. Hinds, M. Jones and A. Beige for discussions.
This work was supported in part by the European Union and the U.K. Engineering and
Physical Sciences Research Council.
\vspace{-0.4cm}


\end{multicols}
\end{document}